\documentclass[prd,tightenlines,nofootinbib,showpacs,preprintnumbers,superscriptaddress, longbibliography]{revtex4}
\usepackage{amsfonts,amsmath,amssymb,amsthm,bbm,hyperref}
\usepackage{graphicx}
\usepackage{color}

\newcommand{\be}{\begin{equation}}
\newcommand{\ee}{\end{equation}}
\newcommand{\beq}{\begin{eqnarray}}
\newcommand{\eeq}{\end{eqnarray}}

\begin{document}

\title{Effects of inter-universal entanglement on the state of the early universe}\thanks{Prepared for the Conference Cosmology on Small Scales 2018, \url{http://css2018.math.cas.cz/}}

\author{Salvador Robles-P\'{e}rez}
\affiliation{Estaci\'{o}n Ecol\'{o}gica de Biocosmolog\'{\i}a, Pedro de Alvarado, 14, 06411 Medell\'{\i}n, Spain.}
\affiliation{IES Miguel Delibes, Miguel Hern\'{a}ndez, 2, 28991 Torrej\'{o}n de la Calzada, Spain.}
\affiliation{Instituto de  F\'{\i}sica Fundamental, CSIC, Serrano 121, 28006 Madrid, Spain}

\begin{abstract}
The creation of universes in entangled pairs with opposite values of the momenta conjugated to the configuration variables of the minisuperspace would be favoured in quantum cosmology by the conservation of the total momentum, in a parallel way as particles are created in pairs with opposite values of their momenta in a quantum field theory. Then, the matter fields that propagate in the two universes may become entangled too, the result of which is the appearance of a quasi thermal state that would produce a specific and distinguishable pattern in the spectrum of fluctuations of the matter fields in the early universe.
\end{abstract}



\maketitle



\section{Introduction}\label{introduction}

The non local effect of quantum entanglement is probably the distinguishing feature of quantum mechanics \cite{Schrodinger1936b} and certainly the one that most departures from the intuition of classical mechanics. In classical mechanics the closer the systems are the stronger are the effects of any interaction between them. This is not necessarily the case in quantum mechanics. Quantum correlations may have a direct effect in the quantum state of one of the interacting systems irrespective of the distance it is separated from the other.

For instance, let us consider the typical example of quantum entanglement consisted in the generation of particles with spin $\frac{1}{2}$. Let $|+\rangle$ and $|-\rangle$ be the corresponding quantum states of the positive and negative spin, respectively, along the $z$ axis. Because the conservation of the spin the particles must be created in pairs in a composite entangled state given by
\be
| \psi \rangle = \frac{1}{\sqrt{2}} \left(  |+\rangle_1 |-\rangle_2 \pm   |-\rangle_1 |+\rangle_2 \right) .
\ee
If we  perform a measurement over one of the particles in the $\left\{|+\rangle, |-\rangle\right\}$ basis, whatever is the result that we obtain, we  know at the same time the state of the other particle, regardless of the distance between them. This is essentially the non-local effect of the quantum entanglement.

\begin{figure}
\centering
\includegraphics[width=10cm]{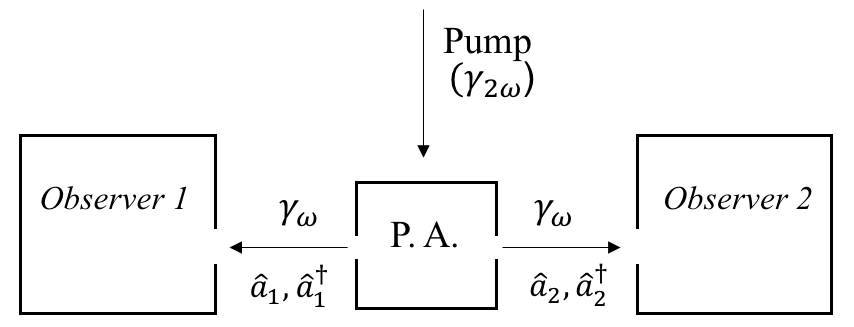}
\caption{Parametric amplifier.}
\label{figure01}
\end{figure}

Another example of quantum entanglement, probably less known but more interesting for the kind of things we are dealing with in this presentation, is the parametric amplifier of quantum optics (see, for instance, Refs. \cite{Walls2008, Scully1997}). The parametric amplifier is a non-linear device that splits a photon of frequency $2\omega$ supplied by a pump into two photons each with frequency $\omega$ (see, Fig. \ref{figure01}). The Hamiltonian of interaction between the classical pump and the two modes $\hat a_1$, $\hat a_1^\dag$ and $\hat a_2$, $\hat a_2^\dag$ is \cite{Walls2008}
\be
H_I = i \hbar \chi \left( \hat a_1^\dag \hat a_2^\dag e^{-2i\omega t} - \hat a_1 \hat a_2 e^{2i\omega t} \right) ,
\ee
where $\chi$ is a constant related to the properties of the non-linear optical medium. The solutions to the Heisenberg equations in the interaction picture are
\beq
\hat a_1(t) &=& \hat a_1 \, \cosh\chi t +  \hat a_2^\dag \, \sinh\chi t   \\ 
\hat a_2(t) &=& \hat a_2 \, \cosh\chi t +  \hat a_1^\dag \, \sinh\chi t ,
\eeq
and the initial two mode vacuum state of the field evolves into a linear combination of perfectly correlated number states \cite{Walls2008}
\be\label{Psi}
|\Psi(t) \rangle = e^{\chi t \left( \hat a_1^\dag \hat a_2^\dag - \hat a_1 \hat a_2 \right) } |0_1\rangle |0_2 \rangle = \frac{1}{\cosh \chi t} \sum_n \left( \tanh\chi t\right)^n |n_1\rangle |n_2\rangle .
\ee
Now, the state of the mode $1$($2$) is given by the reduced density matrix $\rho_{1(2)}$ that is obtained by tracing out from the composite state (\ref{Psi}) the degrees of freedom of the partner mode $2$($1$). It turns out to be \cite{Walls2008}
\be\label{RHO00}
\rho_{1(2)} = {\rm Tr}_{2(1)} \{ |\Psi(t)\rangle \langle \Psi(t)| \} = \frac{1}{Z} \sum_n e^{-\frac{1}{T} (n+\frac{1}{2})} |n_{1(2)}\rangle \langle n_{1(2)}| ,
\ee
where, $Z=\frac{1}{2}\sinh 2\chi t$, and $T^{-1}(t) =  \ln\tanh^{-2}\chi t$. The state (\ref{RHO00}) is a quasi thermal state with a very specific pattern of its thermodynamical properties. If no entanglement were present the modes would remain in their vacuum states and the observers in the boxes (see, Fig. \ref{figure01}) would measure no particle at all in their photo-counters. The interesting thing is then that even if the observers are isolated from each other, if they recognise the pattern (\ref{RHO00}) in the state of their corresponding modes they can infer with high probability of success the existence of the partner modes and the entanglement between them.

The example of the parametric amplifier is specially interesting in the case of cosmology because an expanding spacetime may act as a classical pump for  the modes of the matter fields that propagate along distant regions of the spacetime. Then, analogously to the case of the parametric amplifier, the state of the field in two classically disconnected regions may become represented by a quasi thermal state with a specific pattern of the thermodynamical properties that would depend on the rate of entanglement between the modes of the field in the two regions. The remarkable thing is then that even if the observer may have no direct access to a very distant region of the spacetime, if he or she finds the field in a quasi thermal state with a pattern that follows that of the entangled case, then, he/she can infer the existence of the distant region and the entanglement of the corresponding modes just from the state of the field in the own region!

This is then directly applicable to the case of a multiverse, where by the multiverse one can generally understand the consideration of many different copies of a single piece of the spacetime, each one with their own properties. In principle, all the copies except the one in which we live should be disregarded as being physically inaccessible, and thus, redundant or irrelevant. The most one could do is to take statistical measures of all the universes of the multiverse and formulate statistical predictions. This is what we can call the statistical paradigm of the multiverse (see, for instance, Refs. \cite{Vilenkin1995, Garriga2000, Garriga2001}) and it was the only paradigm of the multiverse taken into account until recent years. However, the power of predictability of the statistical paradigm of the multiverse is quite reduced because in order to make predictions one must assume a principle of mediocrity \cite{Vilenkin1995}, i.e. one has to assume that our universe is one of the most probable universes of the multiverse. Making some sense because it generalises the Copernican principle, the assumption of the mediocrity principle is controversial because we could well be a very strange phenomena of nature violating no physical law. Furthermore, because the statistical paradigm of the multiverse predicts everything within it, then, no concrete prediction can be done actually for a single universe (i.e. everything is possible in the multiverse).

A different paradigm of the multiverse has been developed in the last decade. It can be called the interacting multiverse \cite{Mersini2008c, Mersini2008d, RP2010, RP2011b, Alonso2012}, where entanglement and other non local interactions can be present between the states of the corresponding spacetimes and the fields of the classically disconnected regions. The universes still conserve their classical meaning because they are  isolated from a classical point of view, i.e. light signals cannot joint events of two different universes and therefore no causal relation may exist between their events, at least in the classical (local) sense. Even though, quantum correlations and other non local interactions may still be present, which would be ultimately rooted in the choice of boundary conditions at the origin of the universe or in a residual effect from the underlying theory, whether this can be one of the string theories or the quantum theory of gravity. These non local effects are expected to be large in the earliest stage of the universe and they could propagate and imprint some effect in the observable properties of a large universe like ours. In that case, the multiverse, and the underlying theories, would become a testable proposal as any other theory in cosmology.


\section{Quantum cosmology and the state of the early universe}\label{quantumstate}

Quantum cosmology is the application of the quantum theory to the universe as a whole, i.e. to the spacetime and the matter fields that propagate therein, all together. In the canonical picture the state of the universe is given by a wave function that depends on all the degrees of freedom of the spacetime and the matter fields, and it is the solution of the quantum version of the Hamiltonian constraint,
\be\label{HC01}
\hat H \Psi = 0 ,
\ee
where $\hat H$ is the operator form of the Hamiltonian that corresponds to the Einstein-Hilbert action of the spacetime plus the action of the matter fields. However, for most of the evolution of the universe this can be described by a slow changing background spacetime and small energy fields propagating therein. In that case, the Hamiltonian constraint (\ref{HC01}) can be re-written as
\be\label{HC02}
\left( \hat H_{bg} + \hat H_m \right) \Psi = 0 ,
\ee
where $\hat H_{bg}$ is the Hamiltonian of the background spacetime and $\hat H_m$ is the Hamiltonian of the rapid varying fields. The wave function $\Psi(q_{bg}, q_m)$ depends then on the degrees of freedom of the background spacetime, $q_{bg}$, and on the matter degrees of freedom, $q_m$. In the semiclassical regime, it can be written as a WKB solution of the form \cite{Hartle1990}
\be\label{SCWF01}
\Psi = C(q_{bg}) e^{\pm \frac{i}{\hbar} S_0(q_{bg})} \chi(q_{bg}, q_m) ,
\ee
where $C(q_{bg})$ is a slow varying function of the background variables, $S_0(q_{bg})$ is the action of the background spacetime, and $\chi(q_{bg}, q_m)$ is the wave function of the fields that propagate in the background spacetime. Inserting the semiclassical wave function (\ref{SCWF01}) into the Hamiltonian constraint (\ref{HC02}) and solving it order by order in $\hbar$, one obtains: at zero order in $\hbar$, the classical equations of the background spacetime; and at first order in $\hbar$, the quantum equations of the fields that propagate therein. Thus, the wave function $\Psi$ and the Hamiltonian constraint (\ref{HC02}) contain all the physical information about the classical spacetime  and the quantum matter fields. In that sense, the wave function $\Psi$ represents the state of the whole universe.

For the shake of concreteness, let us consider a homogeneous and isotropic spacetime with a scalar field $\varphi$ propagating therein. Then, the wave function $\Psi$ can be written as the product of two wave functions \cite{Kiefer1987, RP2018a}
\be\label{Psi01}
\Psi(a,\varphi_0 ; x_{n}) = \Psi_0(a,\varphi_0) \chi(a,\varphi_0; x_{n}) .
\ee
The wave function $\Psi_0$ represents the state of the homogeneous and isotropic background, which is entirely described by the dynamics of the scale factor $a$ and the homogeneous mode of the scalar field $\varphi_0$. It is the solution of the Wheeler-DeWitt equation \cite{Kiefer1987, Hartle1990}
\be\label{WDW01}
 \hat H_{bg}  \Psi_0 = \frac{1}{2 a} \left( \hbar^2 \frac{\partial^2}{\partial a^2} + \frac{\hbar^2}{a} \frac{\partial}{\partial a} - \frac{\hbar^2}{a^2} \frac{\partial^2}{\partial \varphi_0^2} + 2 a^4 V(\varphi_0) - a^2  \right)  \Psi_0 = 0 ,
\ee
where $V(\varphi_0)$ is the potential of the scalar field. All the information about the inhomogeneous modes of the scalar field, $x_n$, which are here treated as small perturbations, is encoded in  the wave function $\chi(a,\varphi_0; x_{n})$ in (\ref{Psi01}). It is now easy to show that the wave function $\Psi$ contains: at zero order in $\hbar$, the dynamical information of the classical background spacetime and, at first order in $\hbar$, the quantum information of the inhomogeneous modes $x_n$ that propagate along the background spacetime. Let us first notice that in the semiclassical regime $\Psi_0$ can be written as
\be\label{SCWF03}
\Psi_0(a,\varphi_0) = C(a,\varphi_0) e^{-\frac{i}{\hbar} S(a,\varphi_0)} .
\ee
In that case, the Wheeler-DeWitt equation (\ref{WDW01}) is satisfied at zero order in $\hbar$ if $S(a,\varphi_0)$ is a function that satisfies the Hamilton-Jacobi equation \cite{Kiefer1987}
\be\label{HJ01}
-\left( \frac{\partial S}{\partial a} \right)^2 +\frac{1}{a^2} \left( \frac{\partial S}{\partial \varphi_0} \right)^2 + 2 a^4 V(\varphi_0) - a^2 = 0 .
\ee
Now, choosing as the time variable the WKB parameter $t$ defined by \cite{Kiefer1987} 
\be\label{WKBt01}
\frac{\partial}{\partial t} \equiv  \frac{1}{a} \frac{\partial S}{\partial a}\frac{\partial }{\partial a} - \frac{1}{a^3} \frac{\partial S}{\partial \varphi_0}\frac{\partial }{\partial \varphi_0}  ,
\ee
the equation (\ref{HJ01}) transforms into 
\be\label{FE01}
\dot{a}^2 + 1 - a^2 \left( \dot{\varphi}_0^2 + 2 V(\varphi_0) \right) = 0 ,
\ee
which is the Friedmann equation of the background spacetime. The dynamical equations of $a(t)$ and  $\varphi_0(t)$, given by 
\be\label{MOM01}
\dot{a} =  \frac{1}{a} \left( \frac{\partial S}{\partial a}\right) \ , \ \dot{\varphi_0} = - \frac{1}{a^3} \left( \frac{\partial S}{\partial \varphi_0} \right) ,
\ee
can be directly obtained from (\ref{WKBt01}). Thus, the classical equations of the background spacetime are obtained from the $\hbar^0$ order of the Wheeler-DeWitt equation. On the other hand, inserting the wave function (\ref{Psi01}) into the total Hamiltonian, $H = H_{bg} + H_m$, where $H_m$ is the Hamiltonian of the perturbation modes, it is obtained at first order in $\hbar$ of $H_{bg}$, 
\be\label{SCH00}
i \hbar \left( \frac{1}{a} \frac{\partial S}{\partial a}\frac{\partial }{\partial a} -\frac{1}{a^3} \frac{\partial S}{\partial \varphi_0}\frac{\partial }{\partial \varphi_0} \right) \chi = H_m \chi ,
\ee
which is exactly the Schr\"{o}dinger equation for the inhomogeneous modes of the scalar field if one considers the time variable defined in (\ref{WKBt01}) for the background spacetime. Therefore, the wave function $\Psi$ in (\ref{Psi01}) and the Hamiltonian constraint (\ref{HC02}) contain all the physical information of a single universe. They contain the classical information of the background spacetime and the quantum information of the matter fields that propagate therein.


\section{Creation of universes in entangled pairs}

The symmetries of the Friedmann equation with respect to a time reversal change, $t\rightarrow - t$, in the definition of the time variable (\ref{WKBt01}), and the associated symmetry in the Wheeler-DeWitt equation (\ref{WDW01}) with respect to a change in the sign of the function $S(a,\varphi_0)$, makes that the general solution of the Wheeler-DeWitt equation should be written as
\be\label{Psi02}
\Psi = \sum \Psi^- + \Psi^+ = \sum C_- e^{-\frac{i}{\hbar} S_0} \chi_- + C_+ e^{\frac{i}{\hbar} S_0} \chi_+ ,
\ee
where, $C_+^* = C_-$ and $\chi_+^* = \chi_-$, and the sum extends to all the possible configurations for the semiclassical regime of the spacetime and the matter fields. In (\ref{Psi02}), $\Psi^-$ and $\Psi^+$ are customary referred as the expanding and the contracting branches of the universe, respectively, because in terms of the time parameter $t$ defined in (\ref{WKBt01}) and taking into account the correspondence principle between the classical momentum $p_a^c$ and the quantum momentum, $\hat p_a = - i\hbar\partial_a$, in the classical limit ($\hbar \rightarrow 0$),
\be
- a \dot a \equiv p_a^c \sim \langle \Psi^\pm | \hat p_a | \Psi^\pm\rangle = \pm \frac{\partial S}{\partial a}  ,
\ee
one obtains
\be\label{FE03}
\dot a = \mp \frac{1}{a} \frac{\partial S}{\partial a} ,
\ee
where the $-$ sign corresponds to $\Psi^+$ and the $+$ sign to $\Psi^-$. Then, one should assume that $\Psi^+$ represents a contracting universe and $\Psi^-$ an expanding one. However, the problem with that interpretation is that the time variable $t$ defined in (\ref{WKBt01}) cannot be the physical time in the two branches if for the physical time we mean the time variable measured by a real clock, which is eventually made of matter. The real time is that given in the Schr\"{o}dinger equation, which is the one that ultimately drives the behaviour of matter. If one follows the development of the preceding section with the time defined in (\ref{WKBt01}) for the wave functions $\Psi^+$ and $\Psi^-$ one obtains
\be\label{SCH01}
\mp i \hbar \left( \frac{1}{a} \frac{\partial S}{\partial a}\frac{\partial }{\partial a} -\frac{1}{a^3} \frac{\partial S}{\partial \varphi_0}\frac{\partial }{\partial \varphi_0} \right) \chi_\pm = H_m \chi_\pm ,
\ee
where the $-$ sign corresponds again to $\Psi^+$ and the $+$ sign to $\Psi^-$. Equation (\ref{SCH01}) is the Schr\"{o}dinger equation for the fields $\chi_+$ and $\chi_-$ only if one assumes that the physical time variables in each branch, $t_+$ and $t_-$, respectively, are defined as
\be\label{T01}
t_- = t = - t_+ .
\ee
In that case, each addend in (\ref{Psi02}) corresponds, in terms of their physical time variables, to a pair of both expanding universes that however carry associated opposite values of their momenta conjugated to the configuration variables $a$ and $\varphi_0$. Then, the creation of universes in pairs, as represented in (\ref{Psi02}), entails the conservation of the total momentum conjugated to the configuration variables. Let us notice that the  conservation of the energy is guaranteed even in the case of the creation of a single universe because the gravitational energy is negative and equals the energy of the scalar field. However, the conservation of the generalised momentum $\vec p = (p_a,p_{\varphi_0})$ is only preserved by the creation of universes in pairs with opposite values of their momenta, as it happens in the creation of particles in a quantum field theory.

In fact, the resemblance between the expansion (\ref{Psi02}) and the one made in a quantum field theory is not a coincidence \cite{Strominger1990, RP2010}. The configuration space $\{a,\varphi_0\}$, called the minisuperspace in quantum cosmology \cite{Hartle1990}, can be formally taken as a spacetime with a given geometry and metric element given by \cite{RP2010, RP2018a}
\be\label{MSM01}
d\sigma^2 = - a da^2 + a^3 d\varphi_0^2 .
\ee
From (\ref{MSM01}) one can see that the scale factor formally plays the role of the time like variable of the minisuperspace and the matter field(s) the role of the space like variable(s). The wave function $\Psi(a,\varphi_0)$ can then be seen as a scalar field that \emph{propagates} in the minisuperspace spanned by the variables $\{a,\varphi_0\}$, and thus, a formal parallelism can also be taken between the creation of universes in the minisuperspace and the creation of particles in a curved spacetime. The expansion (\ref{Psi02}) can  be generalised  to
\be\label{PT01}
\Psi(a,\varphi_0) = \int d\mu \left( \Psi^+_\mu \chi_\mu^+ b_\mu + \Psi^-_\mu \chi_\mu^- b^*_\mu \right)  ,
\ee 
where $\mu$ is the set of parameters that determine the transformation (\ref{PT01}), $d\mu$ is the corresponding measure, and $b_\mu$ and $b^*_\mu$ are two constants that can be promoted to the creation and the annihilation operators of universes \cite{RP2017c, RP2018a}. The universes must then be created in pairs with opposite values of their momenta to satisfy  the conservation of the total momentum in the minisuperspace, as particles are created in pairs with opposite values of their momenta in a quantum field theory.

\begin{figure}
\centering
\includegraphics[width=15cm]{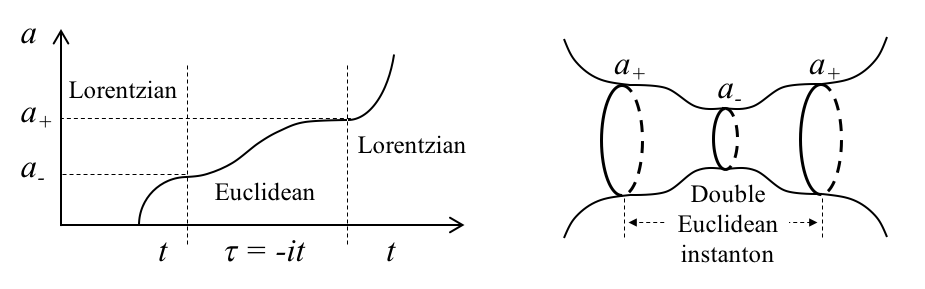}
\caption{Left: there is a Euclidean region between two Lorentzian regions, where classical solutions of the Friedmann equation can be given. Right: a double Euclidean instanton can be formed by matching two single Euclidean instantons, the analytic continuation of which gives rise to two expanding universes.}
\label{figure02}
\end{figure}

For instance, let us consider the inflationary stage of the universe, where the potential of the scalar field can be  considered approximately constant,  $H_0^2 = 2 V(\varphi_0)$, with $V(\varphi_0)$ evaluated at some initial value $\varphi_0(t_0)$. Then, the wave function $\Psi(a,\varphi_0)$ can be expanded in Fourier modes as 
\be\label{WF04}
\Psi(a,\varphi_0) = \int \frac{d K}{\sqrt{2\pi}}  \left( e^{\frac{i}{\hbar} \, K \varphi_0}  \, \Psi^+_K  \,  \chi_K^+  \,  \hat{b}_K 
+  e^{- \frac{i}{\hbar} K \varphi_0}  \,  \Psi^-_K  \, \chi_K^-  \, \hat{b}_K^\dag \right) ,
\ee
where the amplitudes $\Psi^\pm_K(a)$ satisfy
\be\label{WDW02}
\hbar^2 \frac{\partial^2{\Psi^\pm}_K}{\partial a^2} + \frac{\hbar^2}{a} \frac{\partial{\Psi^\pm}_K}{\partial a} + \Omega_K^2(a) \Psi^\pm_K(a) = 0,
\ee
with, 
\be\label{OMEGAK}
\Omega_K = \sqrt{H^2 a^4 - a^2 + \frac{K^2}{a^2}} = \frac{H_0}{a} \sqrt{(a^2 - a_+^2) (a^2-a_-^2) (a^2+a_0^2)} ,
\ee
being $a_+$ and $a_-$ the zeros of the function $\Omega_K$, with $a_+ > a_-$ (see, Refs. \cite{RP2011b, Garay2014}, for the details). The WKB solutions of (\ref{WDW02}) are given by
\be
\Psi_K^\pm \propto \frac{1}{\sqrt{a \Omega_K(a)}} e^{\pm\frac{i}{\hbar}\int \Omega_K(a) da} ,
\ee
and the Friedmann equation of each single universes in terms of their physical times reads [see (\ref{FE03}) with (\ref{T01})]
\be\label{FE02}
\frac{\partial a}{\partial t_\pm} = \frac{1}{a} \Omega_K = \frac{H_0}{a^2} \sqrt{(a^2 - a_+^2) (a^2-a_-^2) (a^2+a_0^2)} .
\ee
There are  two Lorentzian regions for which real solutions of the Friedmann equation (\ref{FE02}) can be obtained, located at $a<a_-$ and at $a>a_+$, respectively. In between there is a Euclidean region that acts as a quantum barrier, where the Euclidean solutions of the Wick rotated version of (\ref{FE02}) are called instantons (see Fig. \ref{figure02}, left). Now, a double Euclidean instanton can be formed by glueing two single instantons at the contact hypersurface $a_-$, whose analytic continuation gives rise to a pair of universes in the Lorentzian region with opposite values of their momentum (see, Fig. \ref{figure02}, right).

Quantum gravitational corrections should  be taken into account as well and they might slightly modify the picture. However, for the case for which the glueing hypersurface $a_-$ is larger enough than the Planck length, the quantum gravitational corrections would be subdominant and the global picture presented here should not be significantly modified.

As explained in Sec. II, the wave functions $\chi_K^+$ and $\chi_K^-$ would  follow the Schr\"{o}dinger equation (\ref{SCH01}) with the time variables $t_+$ and $t_-$ of their respective spacetime backgrounds, i.e.
\be
i\hbar \frac{\partial \chi_\pm}{\partial t_\pm}  = H_m \chi_\pm .
\ee
If we restrict to small linear perturbations the inhomogeneous modes behave like small harmonic oscillators \cite{Halliwell1985, Kiefer1987}. The quantisation of the modes follows as usual, by expanding them in terms of the solutions of the harmonic oscillator and promoting the constants of the expansion into quantum operators,
\be\label{QEx01}
\hat x_{n}(t) = v_n^*(t) \, \hat a_n + v_n(t) \, \hat b_{-n}^\dag ,
\ee
where $v_n(t)$ and $v_n^*(t)$ are two independent solutions of the wave equation
\be\label{HOEQ01}
\ddot{x}_{n} + \frac{3 \dot{a}}{a} \dot{x}_{n} + \omega_n^2 x_{n} = 0 ,
\ee
with a time dependent frequency given by \cite{Kiefer1987, RP2018a}
\be\label{OMEGAM02}
\omega_n^2(t) = \frac{n^2-1}{a^2(t)} - m^2    ,
\ee
where $m$ is the mass of the scalar field.

In the case that the matter content of the entangled universes is represented by a complex scalar field, then, because $\chi_\pm^* = \chi_\mp$, one would expect that the matter is created in the observer's universes and the antimatter in the partner universe \cite{RP2017e}, being both separated by the Euclidean barrier of the double instanton. In that case, $\hat{a}_{n}^\dag$ and $\hat{a}_{n}$ in (\ref{QEx01}) would be the creation and annihilation operators of matter in one of the universes and, $\hat{b}_{n}^\dag$ and $\hat{b}_{n}$,  the creation and annihilation operators of matter in the other one\footnote{Antimatter from the point of view of an observer in the former universe.}. In the case of a real field the particles are their own antiparticles and thus, $\hat a_n = \hat b_n$, with $[\hat a_n, \hat b_n^\dag] = 1$. However, we here retain the different names $\hat{a}_{n}$ and $\hat{b}_{n}$ because in the case of the entangled universes they are still commuting operators as they act on the modes of the field in each single universe of the entangled pair.

We need to impose now the boundary conditions that fix the state of the field. For this, we impose that the perturbation modes are in the composite vacuum state of the invariant representation of the harmonic oscillator (\ref{HOEQ01}). The invariant representation has the great advantage that once the field is in a number state of the invariant representation it remains in the same state along the entire evolution of the field \cite{RP2017d}. In particular, once the field is in the vacuum sate of the invariant representation it remains in the same vacuum state along the entire evolution of the field. It is therefore a stable representation of the vacuum state along the entire evolution of the universes. For the modes $\hat a_n$ and $\hat b_{-n}$, the invariant representation can be written as \cite{Lewis1969, RP2010}
\beq\label{IR01a}
\hat{a}_{n} &=& \sqrt{\frac{1}{2}} \left(  \frac{1}{\sigma} x_{n} + i ( \sigma p_{x_{n}}  - M \dot{\sigma} x_{n}  )  \right) , \\ \label{IR01b}
\hat{b}_{-{n}}^\dag &=& \sqrt{\frac{1}{2}} \left(  \frac{1}{\sigma} x_{n} - i ( \sigma p_{x_{n}}  - M \dot{\sigma} x_{n} )  \right) ,
\eeq
where $\sigma=\sigma(t)$ is an auxiliary function that satisfies a non-linear equation (see, Ref. \cite{RP2017d} and references therein for the details). In that case, the perturbation modes stay in the vacuum state of the invariant representation 
\be\label{VS01}
|0 \rangle = |0_a 0_b \rangle = |0_a\rangle_I  |0_b\rangle_{II} ,
\ee
along the entire evolution of the universes. However, an internal observer would measure the particles of the scalar field in the instantaneous diagonal representation of the harmonic oscillator (\ref{HOEQ01}), which is the representation that defines the instantaneous vacuum state at the moment of the observation. Then, if $\hat c_n$, $\hat c_n^\dag$ and $\hat d_n$, $\hat d_n^\dag$ are the diagonal representations for the universes $I$ and $II$, respectively, they are related to the invariant representation by the Bogolyubov transformation \cite{RP2018a}
\beq\label{BT01a}
{a}_{n} &=& \mu(t) \, {c}_{{n}} - \nu^*(t) \, {d}_{-{n}}^\dag  , \\ \label{BT01b}
{b}_{-{n}} &=&   \mu(t) \, {d}_{-{n}} - \nu^*(t) \, {c}_{{n}}^\dag ,
\eeq
where, $\mu \equiv \mu_n$ and $\nu \equiv \nu_n$, are given by \cite{RP2018a}
\beq\label{MU02}
\mu(t) &=& \frac{1}{2} \left( \sigma \sqrt{a^3 \omega_n} +  \frac{1}{\sigma\sqrt{a^3 \omega_n}}  - i\dot{\sigma} \sqrt{\frac{a^3}{\omega_n }}  \right) , \\ \label{NU02}
\nu(t) &=&\frac{1}{2} \left( \sigma \sqrt{a^3 \omega_n} -  \frac{1}{\sigma\sqrt{a^3 \omega_n}}  - i\dot{\sigma} \sqrt{\frac{a^3}{\omega_n }}  \right) ,
\eeq
with, $|\mu|^2 - |\nu|^2 = 1$ for all time.

The state of the perturbation modes in one single universe of the entangled pair would then be given by the state that is obtained by tracing out from the composite state
\be\label{RHO01}
\rho = | {0}_a {0}_b \rangle \langle {0}_a {0}_b | ,
\ee
the degrees of freedom of the partner universe. Analogously to the example of the parametric amplifier presented in Sec. I, one can show \cite{RP2018a} that the state of the field in each single universe of the entangled pair is 
\be\label{RHO02}
\rho_c = {Tr}_d {\rho}= \prod_{n} \frac{1}{Z_n} \sum_N e^{-\frac{1}{T_n} (N+\frac{1}{2})} | N_{c,{n}} \rangle \langle N_{c,{n}}| ,
\ee
where, $Z_n^{-1} = 2 \sinh\frac{1}{2T_n}$, and, $T_n^{-1}(t)  = \ln\left( 1 + |\nu_n(t)|^{-2}\right).
$
The inhomogeneous modes of the scalar field in each single universe of the entangled pair turn out to be in a quasi-thermal state whose thermal properties depend on the rate of entanglement between the universes. The consideration of the state (\ref{RHO02}) as the initial state of the perturbation modes in the computation of the power spectrum of the CMB would imply a different and possibly distinguishable pattern for the final outcome, so it can provide us with a way of testing definitely the proposal of the creation of universes in entangled pairs.


\section{Observable effects of entangled fields in the early universe}\label{entangledfields}

The observable consequences of the creation of universes in pairs and the subsequent entanglement of the modes of the scalar field that propagate in their spacetimes can be split in two main effects. The first one is the effective modification of the Friedmann equation and, thus, the modification of the evolution of the universes that departures from the inflationary expansion in the very early stage. This pre-inflationary phase of the universe would have observable effects in the power spectrum of the CMB. In particular, it would produce a suppression of the power spectrum of the lowest modes which is currently under investigation \cite{RP2018a, Morais2018}. The other effect is the distribution of the modes of the matter field in a quasi thermal state that should produce specific and distinguishable features in the astronomical data.

In general,  quantum corrections and other non local interactions \cite{Alonso2012, RP2016} are expected to be present in the early stage of the universe and they may induce some observable effect in the  properties of a universe like ours. A first attempt has been made in Ref. \cite{Morais2018}, where it is obtained a suppression of the power spectrum for the lowest modes that could be compatible with the observed data. However, the appearance of an extra peak in that region of the power spectrum invalidates the simplifications made in the model considered there. However, the importance of Ref. \cite{Morais2018} is that it establishes the possibility of observing the effects of quantum cosmology, including those derived from the entanglement between newborn universes, in the properties of the CMB.

Another phenomena that would produce a modification of the Friedmann equation is the backreaction of the matter fields. In our case, the backreaction of the entangled  fields is given by the energy of the modes \cite{RP2018a}
\be\label{ESH01}
\varepsilon =  \frac{H_0^4}{8} \left\{ 1 - \frac{m^2}{H_0^2} \log\frac{b^2}{H_0^2} + \left( 1+ \frac{m^2}{H^2} \right) \left( 1 - \frac{b^2}{H_0^2} \right) \right\} ,
\ee
where $b$ is the SUSY breaking scale of the subjacent landscape \cite{Mersini2008c, Mersini2008d}, and it is expected to produce the same observable imprints to those found in Refs. \cite{Mersini2008c, Mersini2008d, Mersini2017a}. In that case, it would also  produce a suppression of the lowest modes of the CMB. However, a suppression of the lowest modes of the CMB can be produced by many different effects and, furthermore, the dispersion of the observational data in that region of the spectrum is so high that it is not very useful to discriminate between different models. We need therefore a more specific effect to test the creation of the universes in entangled pairs.

The distinguishing feature of our model is that it predicts that the initial state of the field  is in the quasi  thermal state (\ref{RHO02}) that is derived from the entanglement with the modes of a partner universe. It produces a pattern for the spectrum of fluctuations, given in terms of the spectrum of fluctuations of the invariant vacuum (disentangled state) by \cite{RP2018a}
 \be\label{QF01}
 \frac{\delta\phi_\textbf{n}^{th}}{\delta\phi_\textbf{n}^{I}} = \sqrt{\frac{1}{2}\left( 1 + \frac{x^2}{(1+x^2 )(1+\frac{m^2}{H^2 x^2})} \right) }  ,
\ee
that cannot be reproduced by any other known effect, mainly because: i) it is not derived from a vacuum state, and ii) it is not either derived from an exact thermal state because the modes are not thermalised yet in (\ref{RHO02}), i.e. the temperature $T_n$ is not the same for all modes. In fact, one can see that the  large modes ($x\gg 1$) are not affected by the entanglement between the universes. However, the departure from the vacuum state is significant for the horizon modes, $x \sim 1$. That should produce distinguishable  effects in the properties of the CMB and, thus, it might help us to discriminate if our universe was created as a twin universe in an entangled pair.


\section{Conclusions}\label{conclusions}

There is a formal parallelism between the quantum description of the wave function that represents homogeneous and isotropic universes in quantum cosmology and the quantum description of a scalar field that propagates in a curved spacetime. It allows us to consider the creation of the universes in entangled pairs as the most favoured way in which the universes can be created because, only in that case, the total momentum associated to the configuration variables of the minisuperspace is conserved. The two created universes are usually referred as the expanding branch and the contracting branch. However, they are both expanding universes in terms of the time variable that appears in the Schr\"{o}dinger equation of each single universe, which is eventually the physical time variable provided by actual clocks.

The matter fields that propagate in the pair of newborn universes become entangled too, an entanglement that is decreasing along the evolution of the universes but can still be enough to modify the state of the field in the early universe. In particular, the matter field of each single universe becomes represented by a quasi thermal state that would induce a specific and distinguishable pattern in the observable properties of an evolved universe like ours. That makes the interacting multiverse be a testable proposal as any other in cosmology.


\bibliography{../bibliography}

\end{document}